\documentclass[showpacs,aps,prb,superscriptaddress,twocolumn,10pt,amsmath]{revtex4-1}
\usepackage{graphicx}
 
\newcommand{\ket}[1]{\left| #1 \right>}

\begin{document}

\title{Local Kondo temperatures in atomic chains}
\author{R. R. Agundez}
\affiliation{Kavli Institute of Nanoscience, Delft University of Technology, Lorentzweg 1, 2628 CJ Delft, The Netherlands}
\author{J. Salfi}
\affiliation{Centre for Quantum Computation and Communication Technology, University of New South Wales, Sydney NSW 2052, Australia}
\author{S. Rogge}
\affiliation{Kavli Institute of Nanoscience, Delft University of Technology, Lorentzweg 1, 2628 CJ Delft, The Netherlands}
\affiliation{Centre for Quantum Computation and Communication Technology, University of New South Wales, Sydney NSW 2052, Australia}
\author{M. Blaauboer}
\affiliation{Kavli Institute of Nanoscience, Delft University of Technology, Lorentzweg 1, 2628 CJ Delft, The Netherlands}

\begin{abstract}
We study the effect of disorder in strongly interacting small atomic chains. Using the Kotliar-Ruckenstein slave-boson approach we diagonalize the Hamiltonian via scattering matrix theory. We numerically solve the Kondo transmission and the slave-boson parameters that allow us to calculate the Kondo temperature. We demonstrate that in the weak disorder regime, disorder in the energy levels of the dopants induces a non-screened disorder in the Kondo couplings of the atoms. We show that disorder increases the Kondo temperature of a perfect chain. We find that this disorder in the couplings comes from a local distribution of Kondo temperatures along the chain. We propose two experimental setups where the impact of local Kondo temperatures can be observed.
\end{abstract}

\pacs{72.10.Fk, 71.55.Jv, 73.63.-b, 73.63.Kv}

\maketitle

Electron-electron interactions, disorder, and localization are among the most studied phenomena in condensed matter physics. In one-dimensional (1D) systems, disorder and interactions play an important role in transport.  In the limiting case of no electron-electron interactions ($U=0$), we encounter Anderson localization for even the smallest disorder \cite{abrahams_prl42,lee_revmodphys57}. When  interactions dominate, the system is in a Mott insulator state \cite{lee_revmodphys57,abrahams_revmodphys}. Both limiting cases have been extensively studied but much less is known in the intermediate regime. Electron-electron interactions \cite{schubert_physb,miranda_repprogphys,herbut_prb,vettchinka_prb} of intermediate strength can screen disorder, and a metallic phase has been predicted in this regime \cite{book1}. Partial disorder screening gives a gapless two-fluid phase where a fraction of electrons undergo Mott localization and the rest are Anderson localized \cite{book1}.

Recently it has been argued that Kondo physics can play an important role in the conductance of short 1D constrictions. Friedel oscillations are thought to assist the localization of one or more electrons in these short channels. In particular, the so-called $0.7$ anomaly and the zero bias anomaly in quantum point contacts has been attributed to the Kondo effect in these spontaneously localized charges.\cite{iqbal_2013,brun_2014}.

This manuscript presents a theoretical model for investigating the electron-electron interactions and disorder on Kondo transport through a spin chain with intermediate interaction strength connected on its endpoints to two reservoirs. In this intermediate electron-electron interaction strength regime we do not expect the indirect magnetic exchange to play a significant role in Kondo transport. For stronger interactions (weaker couplings) nontrivial competition between the Ruderman-Kittel-Kasuya-Yosida coupling and the Kondo effect is predicted to appear \cite{schwabe_2012}. We show that Coulomb interactions screen the disorder potential for weak disorder and therefore the Kondo transmission increases with increasing interactions. We demonstrate that the screened disorder potential induces a non-screened disorder in the coupling between the electron spin in the atoms and the conduction electrons, the Kondo coupling. This disorder in the Kondo couplings is due to different local Kondo temperatures in the chain. In particular, we find that disorder can enhance local Kondo couplings and overall can increase the Kondo temperature of the system ($T_K$) compared with a perfect spin chain. We also find that the amount of enhancement depends on the position in the chain where disorder is introduced. Motivated by these local Kondo temperatures, we device two schemes to experimentally probe them. In the first, we find that an unbiased but strongly tunnel coupled local probe, (e.g. metallic scanned probe tip), creates an enhancement in the system's Kondo temperature that depends on the probe position. In the second scheme we find that a biased local gate increases the system's Kondo temperature, and that this increment is bigger if the corresponding local Kondo temperature is smaller. With these results we demonstrate that local Kondo couplings and local Kondo temperature can serve as important concepts in Kondo physics of interacting quantum wires.

We propose chains of dopant atoms in silicon for the realization of interacting one-dimensional electronic systems. We envision few dopant systems fabricated with technology for placement of dopants with near-atomic precision by scanning tunneling microscopy (STM), or with few nm precision by ion implantation \cite{Tan:2010gi,Fuechsle:2012bl,Prati:2012ht,Weber:2012gv}. The single dopant Coulomb potentials not only provide reproducible confinement and large charging energies favorable for the realization of interacting one-dimensional electron systems, but the precision placement of dopants is also promising for controlling the coupling strength between elements in a chain of spins.

\begin{figure}[htbp]
\centering
\includegraphics[width=.9\columnwidth]{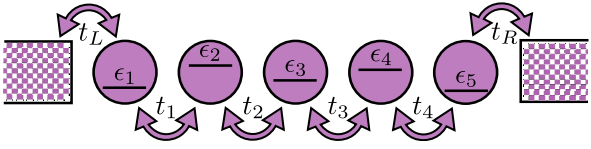}
\caption{Tight-binding model for a chain of QDs coupled to conducting leads.}\label{fig1} 
\end{figure}

\emph{Model}.-
We consider a chain of $N$ dopant atoms connected to conducting leads. We model the structure as a chain of $N$ quantum dots (QDs) whose ends are attached to an electron bath with a dispersion energy given by $\epsilon_k=\epsilon_F-2t_0\cos{k}$. From now on we take $\epsilon_F=0$. Fig. \ref{fig1} shows a schematic representation of the system. We assume that each dopant has a single energy level labeled $\epsilon_i$ with $i=1,2,3,...,N$. The hopping energy between the chain and the left (right) lead is denoted by $t_L$ ($t_R$) and the hopping energy between QD $i$ and QD $i+1$ is $t_i$. Finally, each of the dopants in the chain have an intra-Coulomb interaction $U$.
The Hamiltonian of the system is given by
\begin{eqnarray}\label{hamiltonian}
H=H_{L}+\sum_{i,\sigma}^N{{\epsilon_i f_{i\sigma}^\dagger f_{i\sigma}}}-\hspace{-2mm}\sum_{i}^{N-1}(t_{i}f^\dagger_{i\sigma}f_{i+1\sigma}+c.c.)\nonumber \\
\hspace{-3mm}-\sum_{\sigma}(t_Lc_{-1\sigma}^\dagger f_{1\sigma}+t_Rc_{1\sigma}^\dagger f_{N\sigma}+c.c)+U\sum_i^Nn_{i\uparrow}n_{i\downarrow}
\end{eqnarray}
with $H_L=\epsilon_F\sum_{j\sigma}{c_{j\sigma}^\dagger c_{j\sigma}}-t_0\sum_{j\sigma}(c_{j\sigma}^\dagger c_{j+1\sigma}+c.c.)$ and $n_{i\sigma}=f_{i\sigma}^\dagger f_{i\sigma}$. The operators $c_{j,\sigma} (c_{j,\sigma}^\dagger)$ and $f_{i,\sigma} (f_{i,\sigma}^\dagger)$ correspond to the annihilation (creation) 
operators in the leads and in the QDs, respectively.
Diagonal disorder ($W$) is introduced by randomly choosing $\epsilon_n$ from a uniform distribution centered around the electron-hole symmetry point $-\frac{U}{2}$, that is, $\epsilon_i=-\frac{U}{2}+[-\frac{W}{2},\frac{W}{2}]$, such that for zero disorder the dopants are in the middle of the Coulomb blockade region. We apply the slave-boson approach of Kotliar and Ruckenstein (KR) \cite{kotliar,dong} to treat the electron interaction $U$ in the system. This scheme introduces 4 boson creation (annihilation) operators $e^\dagger_i$ ($e_i$), $p^\dagger_{i\uparrow}$ ($p_{i\uparrow}$), $p^\dagger_{i\downarrow}$ ($p_{i\downarrow}$) and $d^\dagger_i$ ($d_i$) for each QD of the chain. Now the $e^\dagger_i\ket{0}_i$, $p^\dagger_{i\uparrow}f^\dagger_{i\uparrow}\ket{0}_i$, $p^\dagger_{i\downarrow}f^\dagger_{i\downarrow}\ket{0}_i$ and $d^\dagger_i f^\dagger_{i\uparrow}f^\dagger_{i\downarrow}\ket{0}_i$  states represent the empty state, the spin up singly occupied state, the spin down singly occupied state and the doubly occupied state, respectively. $d^\dagger_i d_i$ thus represents the counting operator for the doubly occupied state. This change of basis is applied 
by replacing $f_{i\sigma}\rightarrow f_{i\sigma}z_{i\sigma}$ in the Hamiltonian $H$ [Eq. (\ref{hamiltonian})], where
$z_{i\sigma}=\frac{e^\dagger_i p_{i\sigma}+p^\dagger_{i\tilde{\sigma}}d_i}{\sqrt{1-d^\dagger_i d_i-p^\dagger_{i\sigma}p_{i\sigma}}\sqrt{1-e^\dagger_i e_i-p^\dagger_{i\tilde{\sigma}}p_{\tilde{i\sigma}}}}
$. Two constraints are applied to the bosons, namely the completeness relation: $e_i^\dagger e_i+\sum_\sigma p^\dagger_{i\sigma} p_{i\sigma}+d_i^\dagger d_i=1$, and the correspondence condition between fermions and bosons: $p^\dagger_{n\sigma} p_{n\sigma}+d_n^\dagger d_i=f^\dagger_{i\sigma} f_{i\sigma}$. We add these two constraints to the Hamiltonian [Eq. (\ref{hamiltonian})] by using the Lagrange multipliers $\lambda_i$ and $\gamma_{i\sigma}$. The new Hamiltonian is then $H_{SB}=H+\lambda_i(e_i^\dagger e_i+\sum_\sigma p^\dagger_{i\sigma} p_{i\sigma}+d_i^\dagger d_i-1)+\sum_\sigma\gamma_{i\sigma}(p^\dagger_{i\sigma} p_{i\sigma}+d_i^\dagger d_i-f^\dagger_{i\sigma} f_{i\sigma})$.\\
Following the steps of the KR method the four boson operators are replaced by their corresponding expectation values \cite{kotliar,dong}. We then arrive at the following effective non-interacting Hamiltonian:
\begin{eqnarray}\label{effective_hamiltonian}
H_{eff}=&H_L&+\sum_{i=1}^N{{\tilde{\epsilon_i} f_{i}^\dagger f_{i}}}-\sum_{i=1}^{N-1}(t_iz_iz_{i+1}f^\dagger_{i}f_{i+1}+c.c.)\nonumber \\
&-&(t_Lz_1c_{-1}^\dagger f_{1}+t_Rz_Nc_{1}^\dagger f_{N}+c.c).
\end{eqnarray}
The energy levels in Eq. (\ref{effective_hamiltonian}) have been renormalized to $\tilde{\epsilon_i}=\epsilon_i-\gamma_i$. We analytically diagonalize $H_{eff}$ [Eq. (\ref{effective_hamiltonian})] using scattering theory \cite{orellana}, and together with the KR equations we solve the system self-consistently and obtain the Lagrange multipliers and bosonic expectation values. We use this values in Eq. (\ref{effective_hamiltonian}) and calculate transmission. We also compute the value of the quasiparticle weight $z_i^2=2\frac{(e_i+d_i)^2[1-(e_i^2+d_i^2)]}{1-(e_i^2-d_i^2)^2}$. This is a very important parameter in the description of the impurity problem. In the case of a single impurity, the Kondo temperature in our formalism is given by $T_{K,KR}=z^2 \Gamma$. If $U=0$ the resonance level has the width $\Gamma=\Gamma_L+\Gamma_R$, we find $\Gamma_l=2t_l^2/t_0$, $l=L, R$. To demonstrate the validity of the approach, we show that the KR results for one quantum dot agree with the widely used expression for the Kondo temperature derived by Haldane [Fig. \ref{fig4}(a)] \cite{haldane1978,tsvelick}, and then perform calculations for more complicated systems.

\emph{Results}.- 
Unless stated otherwise we work at zero temperature, take the wide band limit and the symmetric case: $t_i=t_L=t_R=\frac{t_0}{10}=t$.

\begin{figure}[htbp]
\centering
\includegraphics[width=6cm,height=4.5cm]{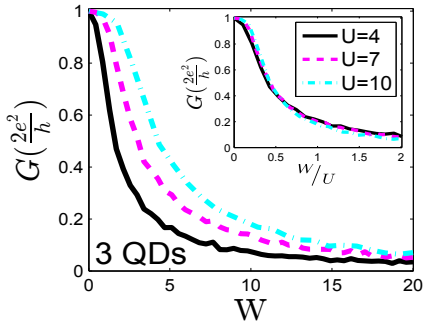}
\caption{Averaged Kondo conductance over 2000 random configurations versus disorder for a chain of $N=3$ QDs for different strengths of electron-electron interaction. The inset shows the curves lying on top of each other when disorder is renormalized by $U$.}\label{fig2} 
\end{figure}

\begin{figure}[htbp]
\centering
\includegraphics[width=7cm,height=6cm]{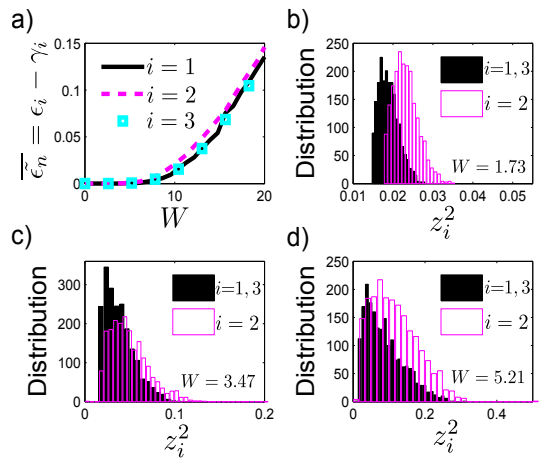}
\caption{Calculations were performed using a chain of 3 QDs with a Coulomb interaction of $U=10$. a) Renormalized energy level of the QDs in the chain averaged over 2000 configurations. b)-d) Distribution of the quasiparticle weight $z_i^2$ for $i=1$ and $i=2$ for all the different system configurations using different values of disorder such that $W<U$.}\label{fig3} 
\end{figure}

\begin{figure}[htbp]
\centering
\includegraphics[width=7cm,height=6.5cm]{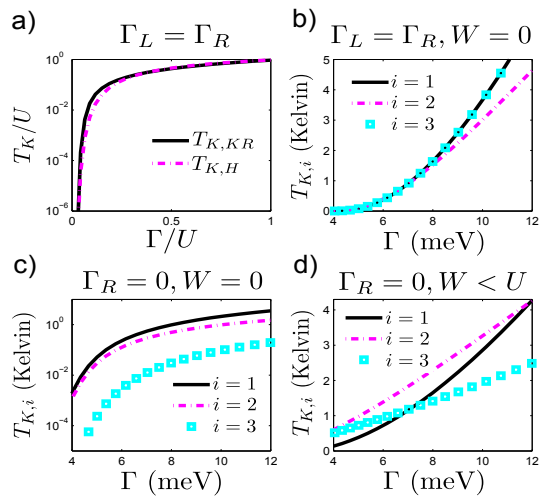}
\caption{a) Comparison of the single impurity Kondo temperature calculated, between the well known relation proposed by Haldane $T_{K,H}=\sqrt{2U\Gamma}e^{-\frac{\pi U}{8\Gamma}}$ (magenta dashed-dot line) and the one for the KR slave-boson method used in our study $T_{K,SB}=z^2\Gamma$ (black full line). b)-d) Local Kondo temperatures reported as $T_{K,i}=z_i^2\Gamma$ for a chain of $N=3$ dopants. b) Symmetric non-disordered system. c) Non-disordered chain attached to one lead. d) Disordered chain attached to one lead. Energy levels were set at $\epsilon_F-\epsilon_1=0.5U$, $\epsilon_F-\epsilon_2=0.2U$ and $\epsilon_F-\epsilon_3=0.8U$.}\label{fig4} 
\end{figure}

We show in Fig. \ref{fig2} averaged values of conductance over 2000 random configurations for different disorder strengths. We observe that if $U$ increases conductance is less sensitive to disorder. We found that in the region $W<U$, the averaged quantum dot levels over the 2000 configurations $\overline{\tilde{\epsilon_i}}$, had values closer to $\epsilon_F$, while for $W>U$ energy levels were not being pinned at the Fermi level anymore [Fig. \ref{fig3}(a)]. This can be attributed to stronger Kondo screening of the disorder for larger values of $U$. For $W<U$ all energy levels lie inside the Coulomb blockade region, therefore all the QDs experience Kondo screening.

As the disorder $W$ increases to the region $W>U$, some of the energy levels can lie outside the Kondo regime (Coulomb blockade region $-U<\epsilon_i<0$). Therefore they will not be pinned to the Fermi level and there is no disorder screening. Since there is no screening by the Coulomb interaction in this region there is no dependence on $U$, so that all the curves merge for $W>>U$ regardless of $U$, as we observe in Fig. \ref{fig2}. For large enough disorder, on-site electron-electron interaction thus plays no role in the Kondo transmission. We would like to point out that our formalism only describes Kondo processes and that in the region $W>U$ there is also transport due to charge fluctuations.

In a plain Hubbard chain the dependence of the conductance on the energy scale $U/t$ can be identified quite easily by dividing the initial Hamiltonian by $t$ such that $U/t$ appears naturally as the energy scale. In this case the interplay of disorder $W$ and Coulomb interaction $U$ is more obscure, and the relevant scale is not naturally found as in the Hubbard case. Moreover the inset of Fig. \ref{fig2} shows how the transmission lines lie on top of each other when disorder is re-normalized by $U$, pointing out that the transmission depends on disorder and Coulomb interaction only via the ratio $W/U$.

In the region $W<U$, disorder is screened by interactions, therefore the energy levels of the dopants are pinned at the Fermi level due to the Kondo effect [Fig. \ref{fig3}(a)]. Despite this fact we observe in Fig. \ref{fig2} a monotonic decrease of transmission for $W<U$. We find that rather than the usual quenching of transmission by the direct diagonal disorder imposed in the Hamiltonian, here disorder in the Kondo couplings quenches the transmission. Diagonal disorder of the applied potential in Eq. (\ref{hamiltonian}) transforms into off-diagonal disorder of couplings in Eq. (\ref{effective_hamiltonian}) via the $z_n$ terms. If disorder increases the values of each $z_i$ can lie in a broader distribution of quasiparticle values [Fig. \ref{fig3}(b)-(d)], and it is the variation of $z_i$ among the sites which creates the monotonic decrease of transmission in the region $W<U$.

We know that the quasiparticle weight can range from 0, signaling the dominance of the Coulomb interaction, to 1, meaning that the spin is behaving freely \cite{first}. For small $W$ the distribution will lie in the lower part of the spectrum. As disorder increases to $W\approx U$ the $z_n$ can be seen to spread and when $W>U$ the distribution will again become narrower, but now around the upper part of the spectrum, signaling the dominance of disorder in the system. We attribute this difference in local Kondo temperatures ($T_{K,i}$) of the impurities in a Kondo chain to the different screening strengths from the conduction electrons in the leads.

\begin{figure}[htbp]
\centering
\includegraphics[width=7cm,height=5cm]{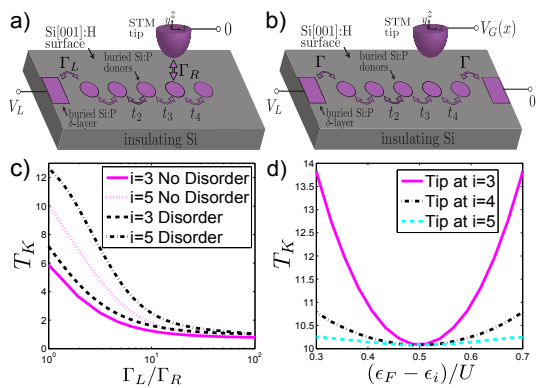}
\caption{a) $T_K$ calculated for a strong coupled lead ($\Gamma_L=7$ meV) and the STM tip that has a tunable coupling to the chain ($\Gamma_R$). b) Calculations for two equally coupled leads and the tip serves as a local gate to the dopants. c) Calculations using setup a), $T_K$ as a function of the degree of symmetry coupling. The tip moves from site $i=3$ to $i=5$ and we model a chain without disorder and one with small disorder were $\epsilon_1\approx-.6U$,  $\epsilon_1\approx-.51U$,  $\epsilon_1\approx-.52U$,  $\epsilon_1\approx-.52U$ and  $\epsilon_1\approx-.63U$. d) Calculations using setup b), $T_K$ versus a single dopant energy level.}\label{fig5} 
\end{figure}

We demonstrate the validity of our slave-boson approach. Fig. \ref{fig4}(a) shows that our results for one quantum dot agree with the widely used expression for the Kondo temperature derived by \citet{haldane1978} for a single Coulomb interacting quantum dot \cite{tsvelick}. For our results presented in Fig. \ref{fig4} and Fig. \ref{fig5} we have used experimental parameters for donors in Silicon \cite{andres_prb,koiller_prl}. The on-site Coulomb repulsion has been estimated as $U\approx 50$ meV, the coupling between the elements of the chain is taken as $t_i=\frac{U}{10}$. This corresponds to a separation of the dopants in silicon of around $7$ nm, which is a realistic experimental value. We have taken $t_0=1.5 U$

We start by calculating the local $T_{K,i}$ for a chain of 3 QDs in a symmetric system with zero disorder [Fig. \ref{fig4}(b)]. This setup simulates an experiment performed in a symmetric arrangement of 3 quantum dots in series, attached to two conducting leads. Our calculations show that even though we get accessible local Kondo temperatures for the system, these temperatures are roughly the same for each QD. For bigger chains we expect this difference to increase. The screening for elements in the middle of the chain will be much weaker than for elements near the end.

We now set $\Gamma_R=0$ to create maximum assymetry in the chain. Conductance will be zero, but the spins are still being screened by the left lead. Setting $\Gamma_R=0$ decreases $T_{K,i}$ in the QDs nearest to the right lead as they will experience less screening from the conduction electrons of the left reservoir. Fig. \ref{fig4}(c) shows the results for zero disorder. We observe that the local Kondo temperatures now vary quite significantly within the chain, and for $\Gamma_L<6$ meV will be challenging to achieve in scanned probe experiments. For an experiment we would like to have Kondo temperatures in the order of a few Kelvins \cite{lansbergen_nanolett}. In Fig. \ref{fig4}(d) we have used a random disorder configuration such that the three dopants are in the Coulomb blockade region. The energy levels of the three dopants are $\epsilon_F-\epsilon_1=0.5U$, $\epsilon_F-\epsilon_2=0.2U$ and $\epsilon_F-\epsilon_3=0.8U$. Our results [Fig. \ref{fig4}(d)] show that disorder can increase the local Kondo temperatures up to a few Kelvins.

We propose that in a chain of atoms the Kondo temperature of the system will be limited by the lowest local Kondo temperature of the chain elements. Next we discuss an experimental setup where coupling to any localized state in the chain can be achieved using an STM operated at low temperature \cite{schofield}. We note that single electron tunneling has been achieved through 3 nm deep dopants having essentially bulk-like properties \cite{salfi}. We introduce two STM experimental setups [Fig. \ref{fig5}(a)-(b)] were all the phenomena in the data can be explained in terms of local Kondo temperatures. Our first setup [Fig. \ref{fig5}(a)] consists of a $N=5$ atomic chain. The left end of the chain is coupled to a left reservoir and the STM tip serves as a right electrode. By variation of the tip-chain separation the coupling ($\Gamma_R$) can be tuned. We calculate $T_K$ by extracting the width of the Kondo resonance in the density of states, we do this by fitting a Lorentzian to the resonance. The results are presented in [Fig. \ref{fig5}(c)]. We can observe that the Kondo temperature for the disordered chain is bigger. Raising local Kondo temperatures by means of disorder (Fig. \ref{fig4}(d)) increases $T_K$ of the system. Also, the influence of the tip is different for different sites. When the tip is over the weakest site, in this case $i=5$ the increment in $T_K$ due to the tip screening is bigger. Finally, when the coupling to the tip is very small the Kondo temperature reaches the same value independently of the tip site, it converges to the $T_K$ of the system left lead-chain.

Another interesting setup consists of two equally coupled leads to the atomic chain, this time the tip serves as a local gate to tune the dopant energy level [Fig. \ref{fig5}(b)]. We calculated the system Kondo temperature in the same way as in the former setup for a non-disordered chain. Fig. \ref{fig5}(d) shows that $T_K $ increases when the energy level of the dopant is displaced from the electron hole symmetry point. By gating a dopant we increase its corresponding local $T_K$, therefore lifting the Kondo temperature of the system. We can also observe that by lifting the lowest local $T_K$ (in this case the atom in the middle $i=3$) we can reach bigger system Kondo temperatures than by gating other atoms. We propose that the identification of local temperatures in a chain of dopants, can be done by a STM measurement of donors in silicon. We expect the system Kondo temperature to be mainly governed by the lowest local $T_K$ in the chain, then by moving along the chain the weakest link  can be identified, proving the existence of local Kondo temperatures in dopant chains.\\
Our observations and calculations lead us to conclude that the Kondo temperature of the system will decrease with the size of the chain. If the chain increases then these new elements will have a lower local $T_K$ hence limiting the system $T_K$. The distribution of local Kondo temperatures due to disorder will spread to lower values for the same reason. We know that parity is an important factor in Kondo, an even number of dopants would not produce a zero bias peak.

\emph{Conclusion}.-
In this manuscript we have shown the existence of induced disorder in the Kondo couplings in an atomic chain due to applied disorder in the energy levels, in the weakly disordered regime ($W<U$). Furthermore we predict different local Kondo temperatures in each of the dopants in the chain and suggest two STM experiments to observe the local Kondo temperatures.

This work is part of the research program of the Foundation for Fundamental Research on Matter (FOM), which is part of the Netherlands Organization for Scientific Research (NWO). This research was financially supported by the ARC Centre of Excellence for Quantum Computation and Communication Technology (CE110001027) and the Future Fellowship (FT100100589).
\bibliography{biblio}

\begin{thebibliography}{26}%
\makeatletter
\providecommand \@ifxundefined [1]{%
 \@ifx{#1\undefined}
}%
\providecommand \@ifnum [1]{%
 \ifnum #1\expandafter \@firstoftwo
 \else \expandafter \@secondoftwo
 \fi
}%
\providecommand \@ifx [1]{%
 \ifx #1\expandafter \@firstoftwo
 \else \expandafter \@secondoftwo
 \fi
}%
\providecommand \natexlab [1]{#1}%
\providecommand \enquote  [1]{``#1''}%
\providecommand \bibnamefont  [1]{#1}%
\providecommand \bibfnamefont [1]{#1}%
\providecommand \citenamefont [1]{#1}%
\providecommand \href@noop [0]{\@secondoftwo}%
\providecommand \href [0]{\begingroup \@sanitize@url \@href}%
\providecommand \@href[1]{\@@startlink{#1}\@@href}%
\providecommand \@@href[1]{\endgroup#1\@@endlink}%
\providecommand \@sanitize@url [0]{\catcode `\\12\catcode `\$12\catcode
  `\&12\catcode `\#12\catcode `\^12\catcode `\_12\catcode `\%12\relax}%
\providecommand \@@startlink[1]{}%
\providecommand \@@endlink[0]{}%
\providecommand \url  [0]{\begingroup\@sanitize@url \@url }%
\providecommand \@url [1]{\endgroup\@href {#1}{\urlprefix }}%
\providecommand \urlprefix  [0]{URL }%
\providecommand \Eprint [0]{\href }%
\providecommand \doibase [0]{http://dx.doi.org/}%
\providecommand \selectlanguage [0]{\@gobble}%
\providecommand \bibinfo  [0]{\@secondoftwo}%
\providecommand \bibfield  [0]{\@secondoftwo}%
\providecommand \translation [1]{[#1]}%
\providecommand \BibitemOpen [0]{}%
\providecommand \bibitemStop [0]{}%
\providecommand \bibitemNoStop [0]{.\EOS\space}%
\providecommand \EOS [0]{\spacefactor3000\relax}%
\providecommand \BibitemShut  [1]{\csname bibitem#1\endcsname}%
\let\auto@bib@innerbib\@empty
\bibitem [{\citenamefont {Abrahams}\ \emph {et~al.}(1979)\citenamefont
  {Abrahams}, \citenamefont {Anderson}, \citenamefont {Licciardello},\ and\
  \citenamefont {Ramakrishnan}}]{abrahams_prl42}%
  \BibitemOpen
  \bibfield  {author} {\bibinfo {author} {\bibfnamefont {E.}~\bibnamefont
  {Abrahams}}, \bibinfo {author} {\bibfnamefont {P.~W.}\ \bibnamefont
  {Anderson}}, \bibinfo {author} {\bibfnamefont {D.~C.}\ \bibnamefont
  {Licciardello}}, \ and\ \bibinfo {author} {\bibfnamefont {T.~V.}\
  \bibnamefont {Ramakrishnan}},\ }\href {\doibase 10.1103/PhysRevLett.42.673}
  {\bibfield  {journal} {\bibinfo  {journal} {Phys. Rev. Lett.}\ }\textbf
  {\bibinfo {volume} {42}},\ \bibinfo {pages} {673} (\bibinfo {year}
  {1979})}\BibitemShut {NoStop}%
\bibitem [{\citenamefont {Lee}\ and\ \citenamefont
  {Ramakrishnan}(1985)}]{lee_revmodphys57}%
  \BibitemOpen
  \bibfield  {author} {\bibinfo {author} {\bibfnamefont {P.~A.}\ \bibnamefont
  {Lee}}\ and\ \bibinfo {author} {\bibfnamefont {T.~V.}\ \bibnamefont
  {Ramakrishnan}},\ }\href {\doibase 10.1103/RevModPhys.57.287} {\bibfield
  {journal} {\bibinfo  {journal} {Rev. Mod. Phys.}\ }\textbf {\bibinfo {volume}
  {57}},\ \bibinfo {pages} {287} (\bibinfo {year} {1985})}\BibitemShut
  {NoStop}%
\bibitem [{\citenamefont {Abrahams}\ \emph {et~al.}(2001)\citenamefont
  {Abrahams}, \citenamefont {Kravchenko},\ and\ \citenamefont
  {Sarachik}}]{abrahams_revmodphys}%
  \BibitemOpen
  \bibfield  {author} {\bibinfo {author} {\bibfnamefont {E.}~\bibnamefont
  {Abrahams}}, \bibinfo {author} {\bibfnamefont {S.~V.}\ \bibnamefont
  {Kravchenko}}, \ and\ \bibinfo {author} {\bibfnamefont {M.~P.}\ \bibnamefont
  {Sarachik}},\ }\href {\doibase 10.1103/RevModPhys.73.251} {\bibfield
  {journal} {\bibinfo  {journal} {Rev. Mod. Phys.}\ }\textbf {\bibinfo {volume}
  {73}},\ \bibinfo {pages} {251} (\bibinfo {year} {2001})}\BibitemShut
  {NoStop}%
\bibitem [{\citenamefont {Schubert}\ \emph {et~al.}(2005)\citenamefont
  {Schubert}, \citenamefont {Weiße},\ and\ \citenamefont
  {Fehske}}]{schubert_physb}%
  \BibitemOpen
  \bibfield  {author} {\bibinfo {author} {\bibfnamefont {G.}~\bibnamefont
  {Schubert}}, \bibinfo {author} {\bibfnamefont {A.}~\bibnamefont {Weiße}}, \
  and\ \bibinfo {author} {\bibfnamefont {H.}~\bibnamefont {Fehske}},\ }\href
  {\doibase http://dx.doi.org/10.1016/j.physb.2005.01.232} {\bibfield
  {journal} {\bibinfo  {journal} {Physica B: Condensed Matter}\ }\textbf
  {\bibinfo {volume} {359-361}},\ \bibinfo {pages} {801 } (\bibinfo {year}
  {2005})}\BibitemShut {NoStop}%
\bibitem [{\citenamefont {Miranda}\ and\ \citenamefont
  {Dobrosavljevi\'{c}}(2005)}]{miranda_repprogphys}%
  \BibitemOpen
  \bibfield  {author} {\bibinfo {author} {\bibfnamefont {E.}~\bibnamefont
  {Miranda}}\ and\ \bibinfo {author} {\bibfnamefont {V.}~\bibnamefont
  {Dobrosavljevi\'{c}}},\ }\href
  {http://stacks.iop.org/0034-4885/68/i=10/a=R02} {\bibfield  {journal}
  {\bibinfo  {journal} {Reports on Progress in Physics}\ }\textbf {\bibinfo
  {volume} {68}},\ \bibinfo {pages} {2337} (\bibinfo {year}
  {2005})}\BibitemShut {NoStop}%
\bibitem [{\citenamefont {Herbut}(2001)}]{herbut_prb}%
  \BibitemOpen
  \bibfield  {author} {\bibinfo {author} {\bibfnamefont {I.~F.}\ \bibnamefont
  {Herbut}},\ }\href {\doibase 10.1103/PhysRevB.63.113102} {\bibfield
  {journal} {\bibinfo  {journal} {Phys. Rev. B}\ }\textbf {\bibinfo {volume}
  {63}},\ \bibinfo {pages} {113102} (\bibinfo {year} {2001})}\BibitemShut
  {NoStop}%
\bibitem [{\citenamefont {Vettchinkina}\ \emph {et~al.}(2013)\citenamefont
  {Vettchinkina}, \citenamefont {Kartsev}, \citenamefont {Karlsson},\ and\
  \citenamefont {Verdozzi}}]{vettchinka_prb}%
  \BibitemOpen
  \bibfield  {author} {\bibinfo {author} {\bibfnamefont {V.}~\bibnamefont
  {Vettchinkina}}, \bibinfo {author} {\bibfnamefont {A.}~\bibnamefont
  {Kartsev}}, \bibinfo {author} {\bibfnamefont {D.}~\bibnamefont {Karlsson}}, \
  and\ \bibinfo {author} {\bibfnamefont {C.}~\bibnamefont {Verdozzi}},\ }\href
  {\doibase 10.1103/PhysRevB.87.115117} {\bibfield  {journal} {\bibinfo
  {journal} {Phys. Rev. B}\ }\textbf {\bibinfo {volume} {87}},\ \bibinfo
  {pages} {115117} (\bibinfo {year} {2013})}\BibitemShut {NoStop}%
\bibitem [{\citenamefont {Dobrosavljevi\'{c}}\ \emph
  {et~al.}(2012)\citenamefont {Dobrosavljevi\'{c}}, \citenamefont {Trivedi},\
  and\ \citenamefont {Valles}}]{book1}%
  \BibitemOpen
  \bibfield  {author} {\bibinfo {author} {\bibfnamefont {V.}~\bibnamefont
  {Dobrosavljevi\'{c}}}, \bibinfo {author} {\bibfnamefont {N.}~\bibnamefont
  {Trivedi}}, \ and\ \bibinfo {author} {\bibfnamefont {J.~M.}\ \bibnamefont
  {Valles}},\ }\href@noop {} {\emph {\bibinfo {title} {Conductor Insulator
  Quantum Phase Transitions}}}\ (\bibinfo  {publisher} {Oxford University
  Press},\ \bibinfo {address} {Oxford},\ \bibinfo {year} {2012})\BibitemShut
  {NoStop}%
\bibitem [{\citenamefont {Iqbal}\ \emph {et~al.}(2013)\citenamefont {Iqbal},
  \citenamefont {Levy}, \citenamefont {Koop}, \citenamefont {Dekker},
  \citenamefont {de~Jong}, \citenamefont {van~der Velde}, \citenamefont
  {Reuter}, \citenamefont {Wieck}, \citenamefont {Aguado}, \citenamefont
  {Meir},\ and\ \citenamefont {van~der Wal}}]{iqbal_2013}%
  \BibitemOpen
  \bibfield  {author} {\bibinfo {author} {\bibfnamefont {M.~J.}\ \bibnamefont
  {Iqbal}}, \bibinfo {author} {\bibfnamefont {R.}~\bibnamefont {Levy}},
  \bibinfo {author} {\bibfnamefont {E.~J.}\ \bibnamefont {Koop}}, \bibinfo
  {author} {\bibfnamefont {J.~B.}\ \bibnamefont {Dekker}}, \bibinfo {author}
  {\bibfnamefont {J.~P.}\ \bibnamefont {de~Jong}}, \bibinfo {author}
  {\bibfnamefont {J.~H.~M.}\ \bibnamefont {van~der Velde}}, \bibinfo {author}
  {\bibfnamefont {D.}~\bibnamefont {Reuter}}, \bibinfo {author} {\bibfnamefont
  {A.~D.}\ \bibnamefont {Wieck}}, \bibinfo {author} {\bibfnamefont
  {R.}~\bibnamefont {Aguado}}, \bibinfo {author} {\bibfnamefont
  {Y.}~\bibnamefont {Meir}}, \ and\ \bibinfo {author} {\bibfnamefont {C.~H.}\
  \bibnamefont {van~der Wal}},\ }\href@noop {} {\bibfield  {journal} {\bibinfo
  {journal} {Nature}\ }\textbf {\bibinfo {volume} {501}} (\bibinfo {year}
  {2013})}\BibitemShut {NoStop}%
\bibitem [{\citenamefont {Brun}\ \emph {et~al.}(2014)\citenamefont {Brun},
  \citenamefont {Martins}, \citenamefont {Faniel}, \citenamefont {Hackens},
  \citenamefont {Bachelier}, \citenamefont {Cavanna}, \citenamefont {Ulysse},
  \citenamefont {Ouerghi}, \citenamefont {Gennser}, \citenamefont {Huant},
  \citenamefont {Bayot}, \citenamefont {Sanquer},\ and\ \citenamefont
  {Sellier}}]{brun_2014}%
  \BibitemOpen
  \bibfield  {author} {\bibinfo {author} {\bibfnamefont {B.}~\bibnamefont
  {Brun}}, \bibinfo {author} {\bibfnamefont {F.}~\bibnamefont {Martins}},
  \bibinfo {author} {\bibfnamefont {S.}~\bibnamefont {Faniel}}, \bibinfo
  {author} {\bibfnamefont {B.}~\bibnamefont {Hackens}}, \bibinfo {author}
  {\bibfnamefont {G.}~\bibnamefont {Bachelier}}, \bibinfo {author}
  {\bibfnamefont {A.}~\bibnamefont {Cavanna}}, \bibinfo {author} {\bibfnamefont
  {C.}~\bibnamefont {Ulysse}}, \bibinfo {author} {\bibfnamefont
  {A.}~\bibnamefont {Ouerghi}}, \bibinfo {author} {\bibfnamefont
  {D.}~\bibnamefont {Gennser}, \bibfnamefont {U.and~Mailly}}, \bibinfo {author}
  {\bibfnamefont {S.}~\bibnamefont {Huant}}, \bibinfo {author} {\bibfnamefont
  {V.}~\bibnamefont {Bayot}}, \bibinfo {author} {\bibfnamefont
  {M.}~\bibnamefont {Sanquer}}, \ and\ \bibinfo {author} {\bibfnamefont
  {H.}~\bibnamefont {Sellier}},\ }\href@noop {} {\bibfield  {journal} {\bibinfo
   {journal} {Nat Commun}\ }\textbf {\bibinfo {volume} {5}} (\bibinfo {year}
  {2014})}\BibitemShut {NoStop}%
\bibitem [{\citenamefont {Schwabe}\ \emph {et~al.}(2012)\citenamefont
  {Schwabe}, \citenamefont {G\"utersloh},\ and\ \citenamefont
  {Potthoff}}]{schwabe_2012}%
  \BibitemOpen
  \bibfield  {author} {\bibinfo {author} {\bibfnamefont {A.}~\bibnamefont
  {Schwabe}}, \bibinfo {author} {\bibfnamefont {D.}~\bibnamefont
  {G\"utersloh}}, \ and\ \bibinfo {author} {\bibfnamefont {M.}~\bibnamefont
  {Potthoff}},\ }\href {\doibase 10.1103/PhysRevLett.109.257202} {\bibfield
  {journal} {\bibinfo  {journal} {Phys. Rev. Lett.}\ }\textbf {\bibinfo
  {volume} {109}},\ \bibinfo {pages} {257202} (\bibinfo {year}
  {2012})}\BibitemShut {NoStop}%
\bibitem [{\citenamefont {Tan}\ \emph {et~al.}(2010)\citenamefont {Tan},
  \citenamefont {Chan}, \citenamefont {M{\"o}tt{\"o}nen}, \citenamefont
  {Morello}, \citenamefont {Yang}, \citenamefont {Donkelaar}, \citenamefont
  {Alves}, \citenamefont {Pirkkalainen}, \citenamefont {Jamieson},
  \citenamefont {Clark},\ and\ \citenamefont {Dzurak}}]{Tan:2010gi}%
  \BibitemOpen
  \bibfield  {author} {\bibinfo {author} {\bibfnamefont {K.~Y.}\ \bibnamefont
  {Tan}}, \bibinfo {author} {\bibfnamefont {K.~W.}\ \bibnamefont {Chan}},
  \bibinfo {author} {\bibfnamefont {M.}~\bibnamefont {M{\"o}tt{\"o}nen}},
  \bibinfo {author} {\bibfnamefont {A.}~\bibnamefont {Morello}}, \bibinfo
  {author} {\bibfnamefont {C.}~\bibnamefont {Yang}}, \bibinfo {author}
  {\bibfnamefont {J.~v.}\ \bibnamefont {Donkelaar}}, \bibinfo {author}
  {\bibfnamefont {A.}~\bibnamefont {Alves}}, \bibinfo {author} {\bibfnamefont
  {J.-M.}\ \bibnamefont {Pirkkalainen}}, \bibinfo {author} {\bibfnamefont
  {D.~N.}\ \bibnamefont {Jamieson}}, \bibinfo {author} {\bibfnamefont {R.~G.}\
  \bibnamefont {Clark}}, \ and\ \bibinfo {author} {\bibfnamefont {A.~S.}\
  \bibnamefont {Dzurak}},\ }\href@noop {} {\bibfield  {journal} {\bibinfo
  {journal} {Nano Lett.}\ }\textbf {\bibinfo {volume} {10}},\ \bibinfo {pages}
  {11} (\bibinfo {year} {2010})}\BibitemShut {NoStop}%
\bibitem [{\citenamefont {Fuechsle}\ \emph {et~al.}(2012)\citenamefont
  {Fuechsle}, \citenamefont {Miwa}, \citenamefont {Mahapatra}, \citenamefont
  {Ryu}, \citenamefont {Lee}, \citenamefont {Warschkow}, \citenamefont
  {Hollenberg}, \citenamefont {Klimeck},\ and\ \citenamefont
  {Simmons}}]{Fuechsle:2012bl}%
  \BibitemOpen
  \bibfield  {author} {\bibinfo {author} {\bibfnamefont {M.}~\bibnamefont
  {Fuechsle}}, \bibinfo {author} {\bibfnamefont {J.~A.}\ \bibnamefont {Miwa}},
  \bibinfo {author} {\bibfnamefont {S.}~\bibnamefont {Mahapatra}}, \bibinfo
  {author} {\bibfnamefont {H.}~\bibnamefont {Ryu}}, \bibinfo {author}
  {\bibfnamefont {S.}~\bibnamefont {Lee}}, \bibinfo {author} {\bibfnamefont
  {O.}~\bibnamefont {Warschkow}}, \bibinfo {author} {\bibfnamefont {L.~C.~L.}\
  \bibnamefont {Hollenberg}}, \bibinfo {author} {\bibfnamefont
  {G.}~\bibnamefont {Klimeck}}, \ and\ \bibinfo {author} {\bibfnamefont
  {M.~Y.}\ \bibnamefont {Simmons}},\ }\href@noop {} {\bibfield  {journal}
  {\bibinfo  {journal} {Nature Nanotech}\ }\textbf {\bibinfo {volume} {7}},\
  \bibinfo {pages} {242} (\bibinfo {year} {2012})}\BibitemShut {NoStop}%
\bibitem [{\citenamefont {Prati}\ \emph {et~al.}(2012)\citenamefont {Prati},
  \citenamefont {Hori}, \citenamefont {Guagliardo}, \citenamefont {Ferrari},\
  and\ \citenamefont {Shinada}}]{Prati:2012ht}%
  \BibitemOpen
  \bibfield  {author} {\bibinfo {author} {\bibfnamefont {E.}~\bibnamefont
  {Prati}}, \bibinfo {author} {\bibfnamefont {M.}~\bibnamefont {Hori}},
  \bibinfo {author} {\bibfnamefont {F.}~\bibnamefont {Guagliardo}}, \bibinfo
  {author} {\bibfnamefont {G.}~\bibnamefont {Ferrari}}, \ and\ \bibinfo
  {author} {\bibfnamefont {T.}~\bibnamefont {Shinada}},\ }\href@noop {}
  {\bibfield  {journal} {\bibinfo  {journal} {Nature Nanotech}\ }\textbf
  {\bibinfo {volume} {7}},\ \bibinfo {pages} {443} (\bibinfo {year}
  {2012})}\BibitemShut {NoStop}%
\bibitem [{\citenamefont {Weber}\ \emph {et~al.}(2012)\citenamefont {Weber},
  \citenamefont {Mahapatra}, \citenamefont {Ryu}, \citenamefont {Lee},
  \citenamefont {Fuhrer}, \citenamefont {Reusch}, \citenamefont {Thompson},
  \citenamefont {Lee}, \citenamefont {Klimeck}, \citenamefont {Hollenberg},\
  and\ \citenamefont {Simmons}}]{Weber:2012gv}%
  \BibitemOpen
  \bibfield  {author} {\bibinfo {author} {\bibfnamefont {B.}~\bibnamefont
  {Weber}}, \bibinfo {author} {\bibfnamefont {S.}~\bibnamefont {Mahapatra}},
  \bibinfo {author} {\bibfnamefont {H.}~\bibnamefont {Ryu}}, \bibinfo {author}
  {\bibfnamefont {S.}~\bibnamefont {Lee}}, \bibinfo {author} {\bibfnamefont
  {A.}~\bibnamefont {Fuhrer}}, \bibinfo {author} {\bibfnamefont {T.~C.~G.}\
  \bibnamefont {Reusch}}, \bibinfo {author} {\bibfnamefont {D.~L.}\
  \bibnamefont {Thompson}}, \bibinfo {author} {\bibfnamefont {W.~C.~T.}\
  \bibnamefont {Lee}}, \bibinfo {author} {\bibfnamefont {G.}~\bibnamefont
  {Klimeck}}, \bibinfo {author} {\bibfnamefont {L.~C.~L.}\ \bibnamefont
  {Hollenberg}}, \ and\ \bibinfo {author} {\bibfnamefont {M.~Y.}\ \bibnamefont
  {Simmons}},\ }\href@noop {} {\bibfield  {journal} {\bibinfo  {journal}
  {Science}\ }\textbf {\bibinfo {volume} {335}},\ \bibinfo {pages} {64}
  (\bibinfo {year} {2012})}\BibitemShut {NoStop}%
\bibitem [{\citenamefont {Kotliar}\ and\ \citenamefont
  {Ruckenstein}(1986)}]{kotliar}%
  \BibitemOpen
  \bibfield  {author} {\bibinfo {author} {\bibfnamefont {G.}~\bibnamefont
  {Kotliar}}\ and\ \bibinfo {author} {\bibfnamefont {A.~E.}\ \bibnamefont
  {Ruckenstein}},\ }\href {\doibase 10.1103/PhysRevLett.57.1362} {\bibfield
  {journal} {\bibinfo  {journal} {Phys. Rev. Lett.}\ }\textbf {\bibinfo
  {volume} {57}},\ \bibinfo {pages} {1362} (\bibinfo {year}
  {1986})}\BibitemShut {NoStop}%
\bibitem [{\citenamefont {Dong}\ and\ \citenamefont {Lei}(2001)}]{dong}%
  \BibitemOpen
  \bibfield  {author} {\bibinfo {author} {\bibfnamefont {B.}~\bibnamefont
  {Dong}}\ and\ \bibinfo {author} {\bibfnamefont {X.~L.}\ \bibnamefont {Lei}},\
  }\href {\doibase 10.1103/PhysRevB.63.235306} {\bibfield  {journal} {\bibinfo
  {journal} {Phys. Rev. B}\ }\textbf {\bibinfo {volume} {63}},\ \bibinfo
  {pages} {235306} (\bibinfo {year} {2001})}\BibitemShut {NoStop}%
\bibitem [{\citenamefont {Orellana}\ \emph {et~al.}(2006)\citenamefont
  {Orellana}, \citenamefont {Lara},\ and\ \citenamefont {Anda}}]{orellana}%
  \BibitemOpen
  \bibfield  {author} {\bibinfo {author} {\bibfnamefont {P.~A.}\ \bibnamefont
  {Orellana}}, \bibinfo {author} {\bibfnamefont {G.~A.}\ \bibnamefont {Lara}},
  \ and\ \bibinfo {author} {\bibfnamefont {E.~V.}\ \bibnamefont {Anda}},\
  }\href {\doibase 10.1103/PhysRevB.74.193315} {\bibfield  {journal} {\bibinfo
  {journal} {Phys. Rev. B}\ }\textbf {\bibinfo {volume} {74}},\ \bibinfo
  {pages} {193315} (\bibinfo {year} {2006})}\BibitemShut {NoStop}%
\bibitem [{\citenamefont {Haldane}(1978)}]{haldane1978}%
  \BibitemOpen
  \bibfield  {author} {\bibinfo {author} {\bibfnamefont {F.~D.~M.}\
  \bibnamefont {Haldane}},\ }\href {\doibase 10.1103/PhysRevLett.40.416}
  {\bibfield  {journal} {\bibinfo  {journal} {Phys. Rev. Lett.}\ }\textbf
  {\bibinfo {volume} {40}},\ \bibinfo {pages} {416} (\bibinfo {year}
  {1978})}\BibitemShut {NoStop}%
\bibitem [{\citenamefont {Tsvelick}\ and\ \citenamefont
  {Wiegmann}(1983)}]{tsvelick}%
  \BibitemOpen
  \bibfield  {author} {\bibinfo {author} {\bibfnamefont {A.}~\bibnamefont
  {Tsvelick}}\ and\ \bibinfo {author} {\bibfnamefont {P.}~\bibnamefont
  {Wiegmann}},\ }\href {\doibase 10.1080/00018738300101581} {\bibfield
  {journal} {\bibinfo  {journal} {Advances in Physics}\ }\textbf {\bibinfo
  {volume} {32}},\ \bibinfo {pages} {453} (\bibinfo {year} {1983})}\BibitemShut
  {NoStop}%
\bibitem [{fir()}]{first}%
  \BibitemOpen
  \href@noop {} {}\bibinfo {note} {If the electrons feel a weak Coulomb
  interaction, the probability for any occupation number of the QD will be
  similar, so $e_i^2=p_{i\sigma}^2=d_i^2\rightarrow\frac{1}{4}$, giving
  $z_i\rightarrow 1$. If there is a strong electron-electron interaction the
  energy level will try to be singly occupied,
  $p_i^2=p_{i\uparrow}^2+p_{i\downarrow}^2\rightarrow 1$. Then $e_i^2$ and
  $d_i^2$ will tend to zero, giving $z_i^2\rightarrow 0$.}\BibitemShut {Stop}%
\bibitem [{\citenamefont {Andres}\ \emph {et~al.}(1981)\citenamefont {Andres},
  \citenamefont {Bhatt}, \citenamefont {Goalwin}, \citenamefont {Rice},\ and\
  \citenamefont {Walstedt}}]{andres_prb}%
  \BibitemOpen
  \bibfield  {author} {\bibinfo {author} {\bibfnamefont {K.}~\bibnamefont
  {Andres}}, \bibinfo {author} {\bibfnamefont {R.~N.}\ \bibnamefont {Bhatt}},
  \bibinfo {author} {\bibfnamefont {P.}~\bibnamefont {Goalwin}}, \bibinfo
  {author} {\bibfnamefont {T.~M.}\ \bibnamefont {Rice}}, \ and\ \bibinfo
  {author} {\bibfnamefont {R.~E.}\ \bibnamefont {Walstedt}},\ }\href {\doibase
  10.1103/PhysRevB.24.244} {\bibfield  {journal} {\bibinfo  {journal} {Phys.
  Rev. B}\ }\textbf {\bibinfo {volume} {24}},\ \bibinfo {pages} {244} (\bibinfo
  {year} {1981})}\BibitemShut {NoStop}%
\bibitem [{\citenamefont {Koiller}\ \emph {et~al.}(2001)\citenamefont
  {Koiller}, \citenamefont {Hu},\ and\ \citenamefont
  {Das~Sarma}}]{koiller_prl}%
  \BibitemOpen
  \bibfield  {author} {\bibinfo {author} {\bibfnamefont {B.}~\bibnamefont
  {Koiller}}, \bibinfo {author} {\bibfnamefont {X.}~\bibnamefont {Hu}}, \ and\
  \bibinfo {author} {\bibfnamefont {S.}~\bibnamefont {Das~Sarma}},\ }\href
  {\doibase 10.1103/PhysRevLett.88.027903} {\bibfield  {journal} {\bibinfo
  {journal} {Phys. Rev. Lett.}\ }\textbf {\bibinfo {volume} {88}},\ \bibinfo
  {pages} {027903} (\bibinfo {year} {2001})}\BibitemShut {NoStop}%
\bibitem [{\citenamefont {Lansbergen}\ \emph {et~al.}(2010)\citenamefont
  {Lansbergen}, \citenamefont {Tettamanzi}, \citenamefont {Verduijn},
  \citenamefont {Collaert}, \citenamefont {Biesemans}, \citenamefont
  {Blaauboer},\ and\ \citenamefont {Rogge}}]{lansbergen_nanolett}%
  \BibitemOpen
  \bibfield  {author} {\bibinfo {author} {\bibfnamefont {G.~P.}\ \bibnamefont
  {Lansbergen}}, \bibinfo {author} {\bibfnamefont {G.~C.}\ \bibnamefont
  {Tettamanzi}}, \bibinfo {author} {\bibfnamefont {J.}~\bibnamefont
  {Verduijn}}, \bibinfo {author} {\bibfnamefont {N.}~\bibnamefont {Collaert}},
  \bibinfo {author} {\bibfnamefont {S.}~\bibnamefont {Biesemans}}, \bibinfo
  {author} {\bibfnamefont {M.}~\bibnamefont {Blaauboer}}, \ and\ \bibinfo
  {author} {\bibfnamefont {S.}~\bibnamefont {Rogge}},\ }\href {\doibase
  10.1021/nl9031132} {\bibfield  {journal} {\bibinfo  {journal} {Nano Letters}\
  }\textbf {\bibinfo {volume} {10}},\ \bibinfo {pages} {455} (\bibinfo {year}
  {2010})},\ \bibinfo {note} {pMID: 20041698}\BibitemShut {NoStop}%
\bibitem [{\citenamefont {Schofield}\ \emph {et~al.}(2013)\citenamefont
  {Schofield}, \citenamefont {Studer}, \citenamefont {Hirjibehedin},
  \citenamefont {Curson}, \citenamefont {Aeppli},\ and\ \citenamefont
  {Bowler}}]{schofield}%
  \BibitemOpen
  \bibfield  {author} {\bibinfo {author} {\bibfnamefont {S.~R.}\ \bibnamefont
  {Schofield}}, \bibinfo {author} {\bibfnamefont {P.}~\bibnamefont {Studer}},
  \bibinfo {author} {\bibfnamefont {C.~F.}\ \bibnamefont {Hirjibehedin}},
  \bibinfo {author} {\bibfnamefont {N.~J.}\ \bibnamefont {Curson}}, \bibinfo
  {author} {\bibfnamefont {G.}~\bibnamefont {Aeppli}}, \ and\ \bibinfo {author}
  {\bibfnamefont {D.~R.}\ \bibnamefont {Bowler}},\ }\href {\doibase
  10.1038/ncomms2679} {\bibfield  {journal} {\bibinfo  {journal} {Nature
  Commun.}\ }\textbf {\bibinfo {volume} {4}},\ \bibinfo {pages} {1649}
  (\bibinfo {year} {2013})}\BibitemShut {NoStop}%
\bibitem [{\citenamefont {Salfi}\ \emph {et~al.}(2014)\citenamefont {Salfi},
  \citenamefont {Mol}, \citenamefont {Rahman}, \citenamefont {Klimeck},
  \citenamefont {Simmons}, \citenamefont {Hollenberg},\ and\ \citenamefont
  {Rogge}}]{salfi}%
  \BibitemOpen
  \bibfield  {author} {\bibinfo {author} {\bibfnamefont {J.}~\bibnamefont
  {Salfi}}, \bibinfo {author} {\bibfnamefont {J.~A.}\ \bibnamefont {Mol}},
  \bibinfo {author} {\bibfnamefont {R.}~\bibnamefont {Rahman}}, \bibinfo
  {author} {\bibfnamefont {G.}~\bibnamefont {Klimeck}}, \bibinfo {author}
  {\bibfnamefont {M.~Y.}\ \bibnamefont {Simmons}}, \bibinfo {author}
  {\bibfnamefont {L.~C.~L.}\ \bibnamefont {Hollenberg}}, \ and\ \bibinfo
  {author} {\bibfnamefont {S.}~\bibnamefont {Rogge}},\ }\href {\doibase
  10.1038/nmat3941} {\bibfield  {journal} {\bibinfo  {journal} {Nature Mater.}\
  }\textbf {\bibinfo {volume} {13}},\ \bibinfo {pages} {605} (\bibinfo {year}
  {2014})}\BibitemShut {NoStop}%
\end{thebibliography}%
\bibliographystyle{apsrev4-1}

\end{document}